\documentclass{emulateapj}
\newcommand{\pg}{PG\,1115+080}

\newcommand{\hubble}{\textit{Hubble Space Telescope}}
\newcommand{\hst}{\textit{HST}}
\newcommand{\chandra}{\textit{Chandra}}
\newcommand{\xmm}{\textit{XMM}}
\newcommand{\xmmn}{\textit{XMM-Newton}}
\newcommand{\rosat}{\textit{ROSAT}}
\newcommand{\ergcms}{\ensuremath{\mathrm{erg~cm}^{-2}~\mathrm{s}^{-1}}}
\newcommand{\Lx}{\ensuremath{L_\mathrm{x}}}

\begin{document}
\shorttitle{Flux Ratio Anomalies in \pg}
\shortauthors{Pooley et al.}
\slugcomment{accepted to ApJ}
\title{A Strong X-Ray Flux Ratio Anomaly in the Quadruply Lensed Quasar \pg$^1$}

\author{David Pooley\altaffilmark{2,3}, Jeffrey A.~Blackburne\altaffilmark{4}, Saul~Rappaport\altaffilmark{4}, Paul L.~Schechter\altaffilmark{4}, \& Wen-fai Fong\altaffilmark{3}}

\altaffiltext{1}{Based on observations obtained with the Magellan Consortium's Baade and Clay Telescopes.}
\altaffiltext{2}{University of California at Berkeley, Astronomy Department, 601 Campbell Hall, Berkeley, CA 94720; dave@astron.berkeley.edu}
\altaffiltext{3}{Chandra Fellow}
\altaffiltext{4}{Massachusetts Institute of Technology, Department of Physics and Kavli Institute for Astrophysics and Space Research, 70 Vassar St., Cambridge, MA 02139; jeffb@space.mit.edu, sar@mit.edu, schech@achernar.mit.edu, wenfs@mit.edu}
\begin{abstract}

\pg\ is a quadruply lensed quasar at $z=1.72$ whose image positions are well fit by simple models of the lens galaxy (at $z=0.31$).  At optical wavelengths, the bright close pair of images exhibits a modest flux ratio anomaly (factors of $\sim$1.2--1.4 over the past 22 years) with respect to these same models.  We show here that as observed in X-rays with \chandra, the flux ratio anomaly is far more extreme, roughly a factor of 6.  The contrasting flux ratio anomalies in the optical and X-ray band confirm the microlensing hypothesis and set a lower limit on the size of the optical continuum emission region that is $\sim$10--100 times larger than expected from a thin accretion disk model.

\end{abstract}

\keywords{ gravitational lensing --- quasars: individual (\pg) }

\section{Introduction}
\label{sec:intro}

\pg\ was the second gravitationally lensed quasar to be discovered, and the first found to be quadruple \citep{1980Natur.285..641W}. It has been the subject of numerous studies at wavelengths ranging from radio to mid-infrared to optical to UV to X-ray.  The lensing galaxy is a member of a small group of galaxies, the tide from which produces the quadrupole moment needed to produce four images \citep{1997AJ....114..507K}.  It was the first gravitationally lensed system to yield multiple time delays \citep{1997ApJ...475L..85S}.  The optical images show uncorrelated flux variations on a timescale of order one year, presumably the result of microlensing by stars in the lensing galaxy \citep{1985A&A...149L..13F}.

\pg\ is an example of what \citet{2003AJ....125.2769S} call an ``inclined quad,'' a system with a close, bright pair of images that results when a lensed source lies just inside a ``fold'' caustic \citep{2005ApJ...635...35K}.  Several very similar systems have subsequently been discovered \citep{1992AJ....104..968H, 2004AJ....127.2617M, 2002A&A...382L..26R, 2003AJ....126..666I}. 
In each case, one of the two close images is a minimum of the light travel time surface, and the other is a saddlepoint.  From quite general considerations, if the gravitational potential is smooth, one expects the close, bright pair to be mirror images of each other and therefore very nearly equal in brightness \citep{2002ApJ...567L...5M}. {\em All} of the known inclined quads violate this prediction, despite the fact that such models fit the observed image positions to within a few percent.  This phenomenon has come to be known as the ``flux ratio anomaly'' problem.

In this regard \pg\ is the {\em least} anomalous among the inclined quads.  In the earliest images that resolved the close pair, the ratio of the flux of the saddlepoint ($A_2$) to that of the minimum ($A_1$) was very nearly unity \citep{1986A&A...158L...5V}. By the mid-1980s the ratio had decreased to $\sim$$2/3$.  Recent optical observations (see \S\ref{sec:opt-obs}) give a ratio closer to $\sim$$5/6$.  By contrast the corresponding optical ratios for inclined quads WFI\,J2026$-$4536, HS\,0810+2554, MG\,0414+0534, and SDSS\,J0924+0219 are approximately $3/4$, $1/2$, $1/3$, and $1/10$, respectively \citep{2004AJ....127.2617M,2002A&A...382L..26R,1993AJ....105....1S,2003AJ....126..666I}. For this last case, \citet{2006ApJ...639....1K} argue that microlensing by stars (rather than millilensing by dark matter subcondensations) is responsible for the anomaly.  \citet{2004AAS...205.2806P} and \citet{2006astro.ph..1523M} predict a substantial brightening of the faint saddlepoint in SDSS\,J0924+0219 on a timescale of roughly one decade if the microlensing hypothesis is correct.

\citet{2004AAS...205.2806P} argue that the saddlepoint in \pg\ would likewise be expected to get substantially (a factor of 2 or more) fainter on a similar timescale.  But over the course of a quarter century \pg\ has declined to cooperate, at least at optical wavelengths.

In the present paper we report that \pg\ has indeed been exhibiting microlensing of the expected amplitude, but at X-ray wavelengths rather than at optical wavelengths. In \S\ref{sec:obs} we describe the X-ray and optical observations and our analysis. In \S\ref{sec:discuss} we discuss implications for the lensing galaxy and for the relative sizes of the quasar's optical and X-ray emitting regions. We summarize our conclusions in \S\ref{sec:conclusions}. Throughout, we assume a ``concordance'' cosmology with $\Omega_M=0.3$, $\Omega_{\Lambda}=0.7$, and $h=0.72$.

\setlength{\fboxsep}{0cm}
\setlength{\fboxrule}{1pt}
\begin{figure*}                                                                 
\centering
\framebox{\includegraphics[width=0.46\textwidth]{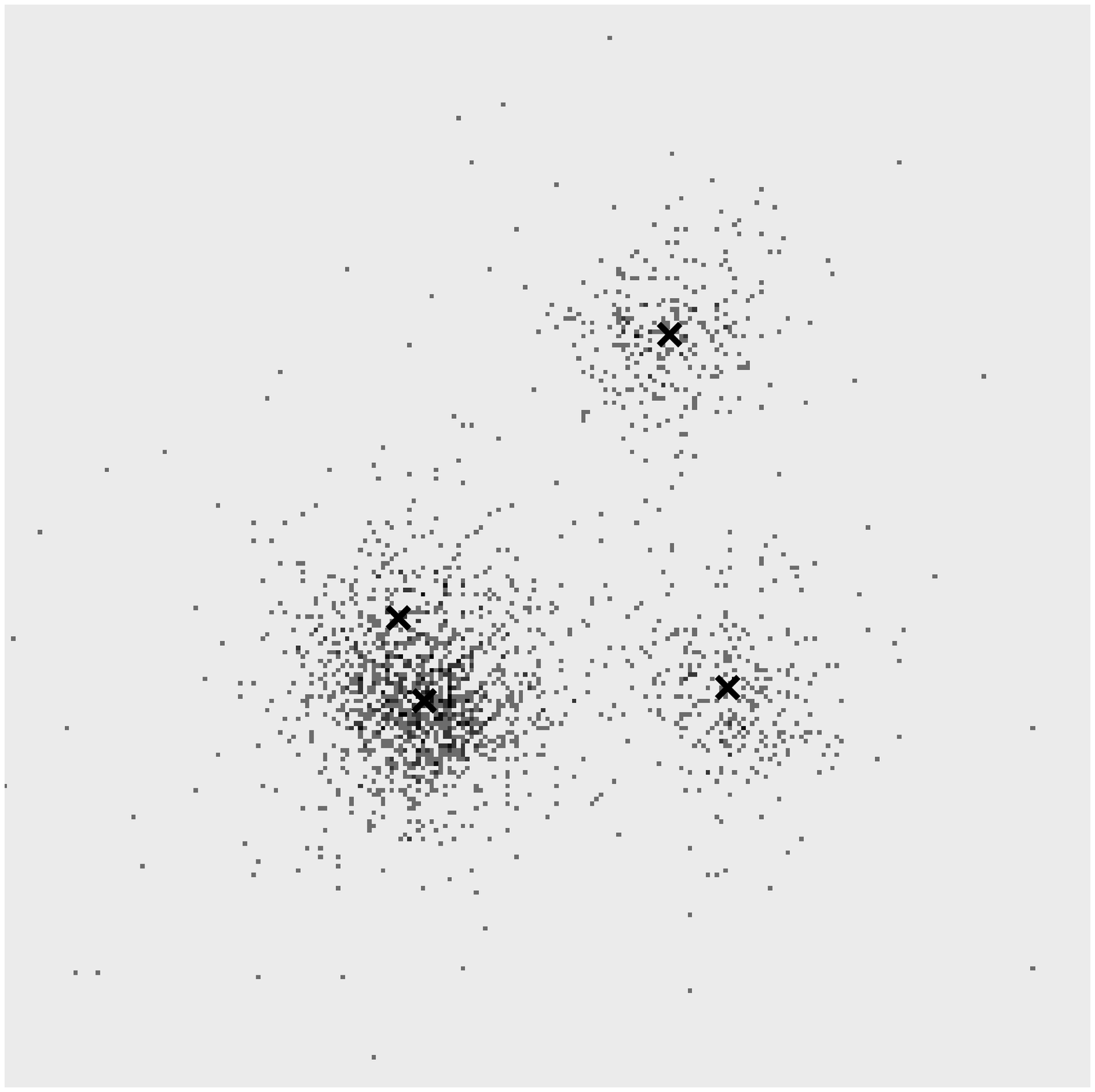}}\hglue0.4cm
\framebox{\includegraphics[width=0.46\textwidth]{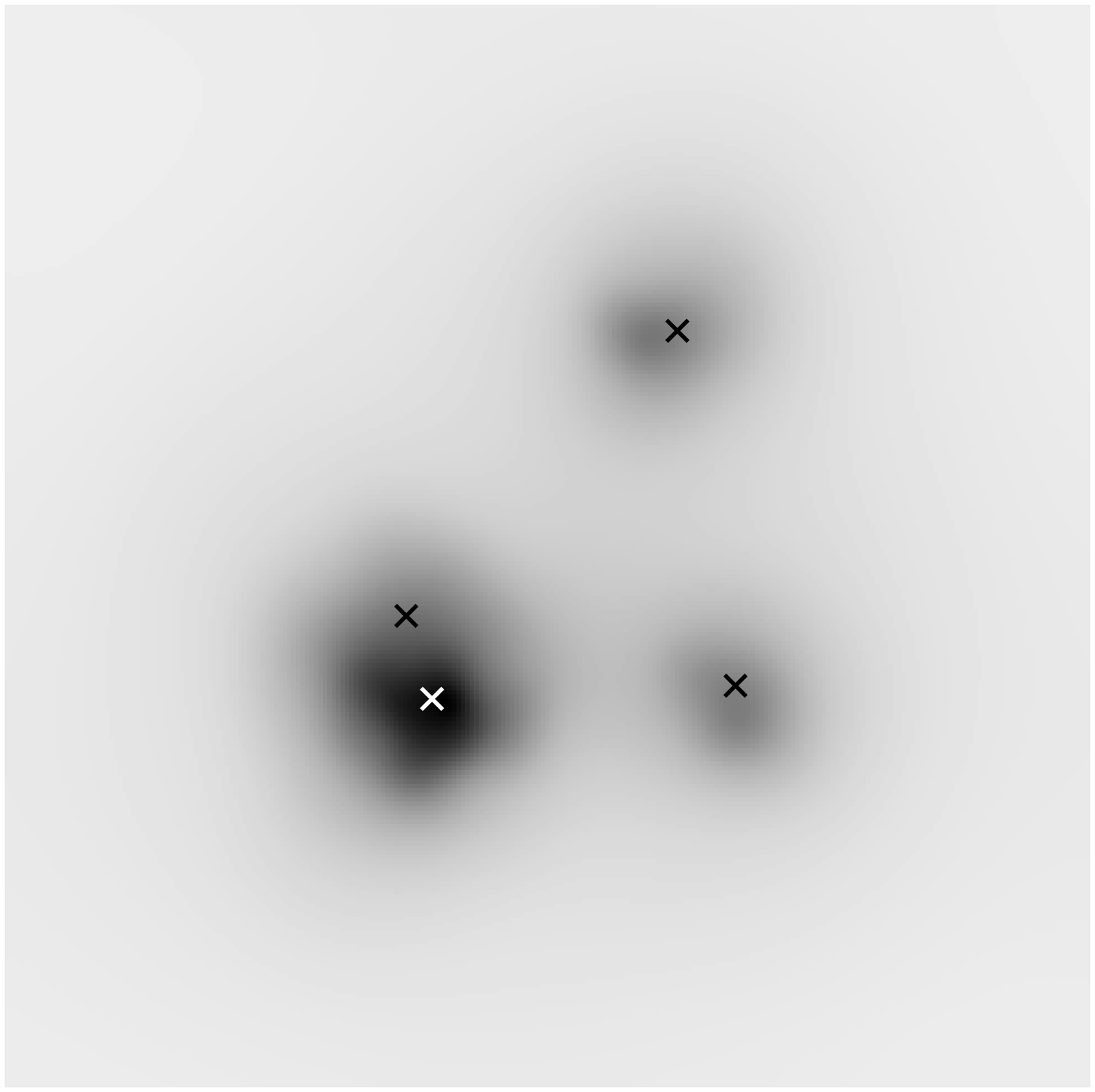}}
\vglue0.3cm
\framebox{\includegraphics[width=0.46\textwidth]{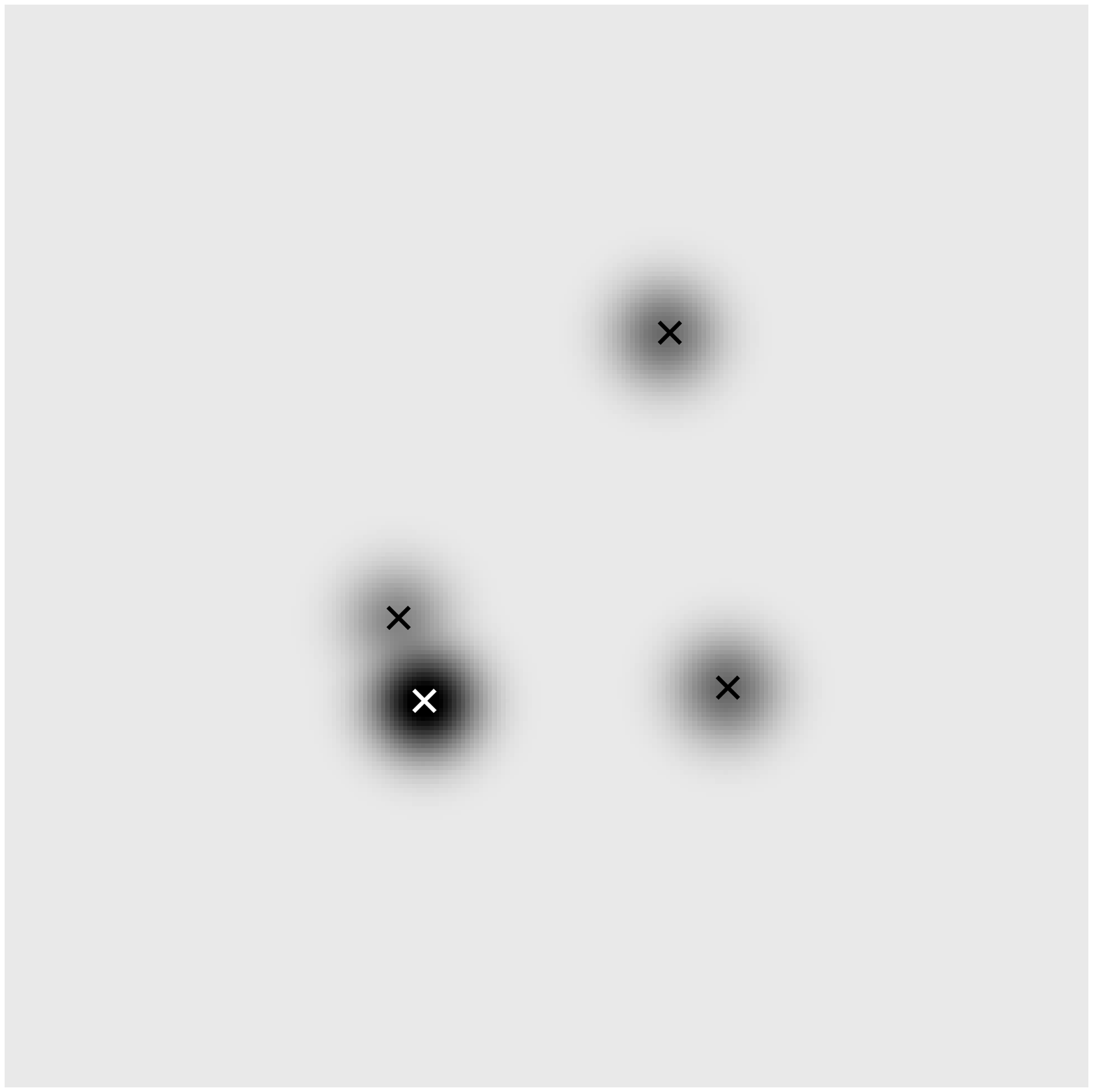}}\hglue0.4cm
\framebox{\includegraphics[width=0.46\textwidth]{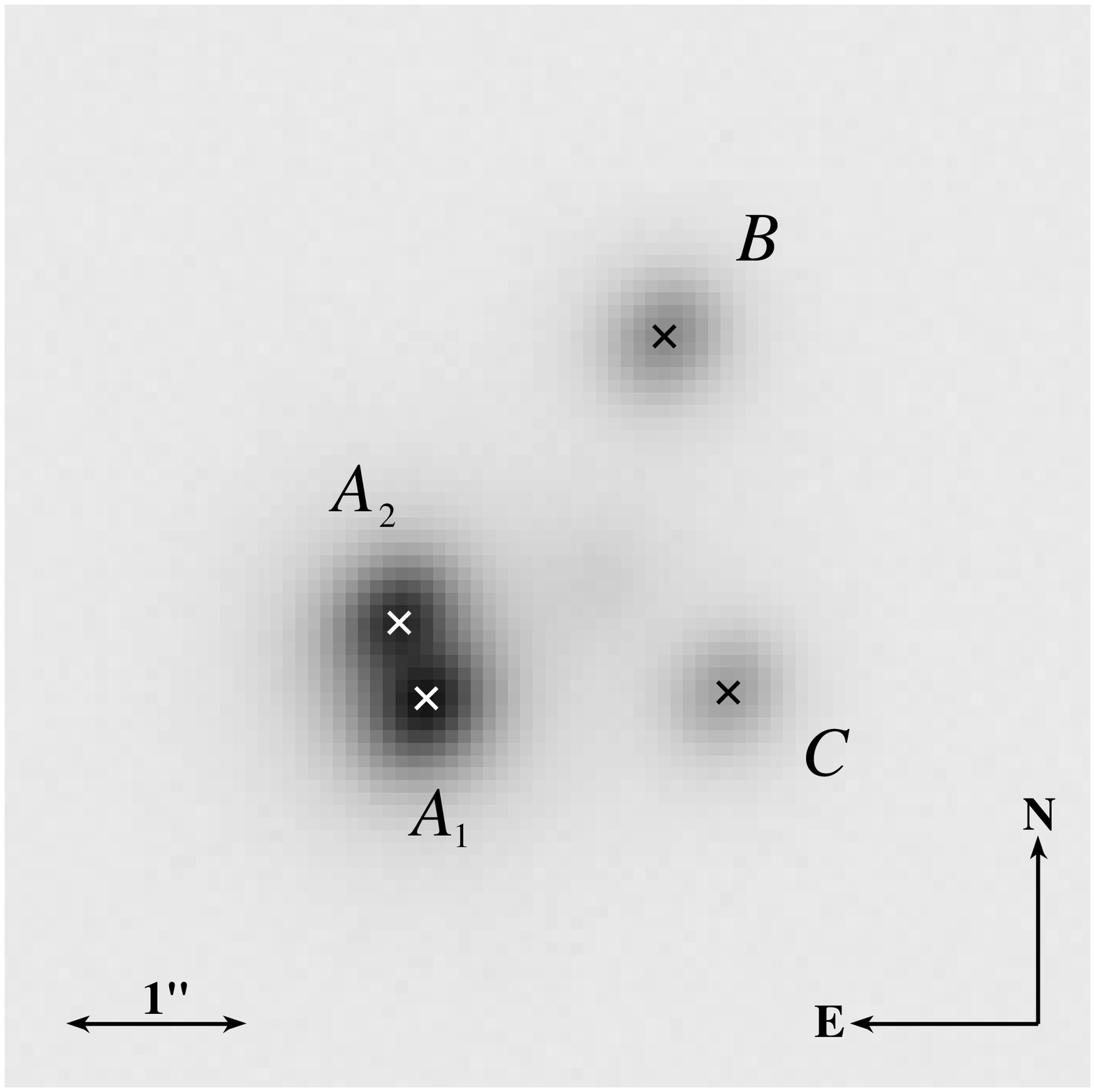}}
\caption{X-ray and optical images of \pg.  Each image is $6''\times6''$. {\it Top left:} Raw \chandra\ data from ObsID 363 (2000 Jun 02) in the 0.5--8 keV band. {\it Top right:} Adaptively smoothed \chandra\ image. {\it Bottom left:} ``Pseudo MEM'' image based on fits to the raw \chandra\ data.  {\it Bottom right:} Sloan $i'$-band Magellan image from 2005 Jun 07.}
\label{fig:images}
\end{figure*}


\section{Observations and Analysis}
\label{sec:obs}

\subsection{X-ray observations}
\label{sec:xray-obs}

\pg\ was observed for 26.5 ks on 2000 Jun 02 (ObsID 363) and for 9.8 ks on 2000 Nov 03 (ObsID 1630) with the \textit{Chandra X-ray Observatory's} Advanced CCD Imaging Spectrometer (ACIS).  These observations were used by \citet{2004ApJ...610..686G} to study the X-ray properties of the lensing group of galaxies.  The data were taken in timed-exposure mode with an integration time of 3.24 s per frame, and the telescope aimpoint was on the back-side illuminated S3 chip. The data were telemetered to the ground in faint mode. 

The data were downloaded from the \chandra\ archive, and reduction was performed using the CIAO\,3.3 software provided by the \chandra\ X-ray Center\footnote{\url{http://asc.harvard.edu}}.  The data were reprocessed using the CALDB\,3.2.1 set of calibration files (gain maps, quantum efficiency, quantum efficiency uniformity, effective area) including a new bad pixel list made with the {\tt acis\_run\_hotpix} tool.  The reprocessing was done without including the pixel randomization that is added during standard processing.  This omission slightly improves the point spread function.  The data were filtered using the standard {\it ASCA} grades and excluding both bad pixels and software-flagged cosmic ray events. Intervals of strong background flaring were searched for, but none were found.

For each observation, an image was produced in the 0.5--8 keV band with a resolution of 0\farcs0246 per pixel (see Figure~\ref{fig:images}).  To determine the intensities of each lensed quasar image, a two-dimensional model consisting of four Gaussian components plus a constant background was fit to the data.  The background component was fixed to a value determined from a source-free region near the lens.  The relative positions of the Gaussian components were fixed to the separations determined from \hubble\ observations \citep{1993AJ....106.1330K}, but the absolute position was allowed to vary.  Each Gaussian was constrained to have the same full-width at half-maximum, but this value was allowed to float.  The fits were performed in Sherpa \citep{2001SPIE.4477...76F} using Cash statistics \citep{1979ApJ...228..939C} and the Powell minimization method.  The intensity ratios (relative to image C) are listed in Table~\ref{tab:xrayfits}.  The best-fit full-width at half maximum (fwhm) was $0\farcs83 \pm 0\farcs01$ for ObsID 363 and $0\farcs80 \pm 0\farcs02$ for ObsID 1630; both consistent with the overall width of the instrumental point spread function (PSF) as found in the \chandra\ PSF Library \citep{2001ASPC..238..435K} supplied by the \chandra\ X-ray Center. In addition to the Gaussians, models of the form $f(r)=A[1+(r/r_0)^2]^{-\alpha}$ were also tried; these gave similar results to the values in Table~\ref{tab:xrayfits}.  

Based on the best fit Gaussian shape and the relative intensities, we constructed a ``pseudo'' maximum entropy method (MEM) representation of the data.  Here we have simply plotted Gaussians of a common width (fwhm = $0.22''$), with the fitted intensities and at the fitted locations (see Table~\ref{tab:xrayfits}).  We used the largest source width consistent (at 3\,$\sigma$ confidence) with no blurring of the intrinsic \chandra\ PSF.  A maximum likelihood deconvolution of the image is presented by \citet{2004mmu..sympE...1C} and appears consistent with our ``pseudo'' MEM image.

\begin{deluxetable}{rccl}
\tablewidth{0pt}
\tablecaption{X-ray and Model Flux Ratios \label{tab:xrayfits}}
\tablehead{
\colhead{Ratio} & \colhead{ObsID 363} & \colhead{ObsID 1630}  & \colhead{Model}}\\
\startdata
$A_1/C$ & $3.9\pm0.3$ & $4.3\pm0.5$ & 3.91 \\
$A_2/C$ & $0.6\pm0.1$ & $1.2\pm0.3$ & 3.73 \\
$B/C$   & $1.0\pm0.1$ & $0.9\pm0.1$ & 0.67 \\
$A_2/A_1$& $0.16\pm0.03$ & $0.29\pm0.08$ & 0.96
\enddata
\end{deluxetable}

Spectra of the quasar images were extracted using the {\tt ACIS Extract} package v3.94 \citep{Broos02}.  A single spectrum of $A_1$ and $A_2$ was extracted because of the significant overlap, but $B$ and $C$ were extracted separately.  Both the \chandra\ effective area and PSF are functions of energy, and {\tt ACIS Extract} corrected the effective area response for each spectrum based on the fraction of the PSF enclosed by the extraction region (at 1.5 keV, these fractions were 0.9 for $A_1+A_2$, 0.8 for $B$, and 0.9 for $C$).  The spectra were grouped to contain at least ten counts per bin, and $\chi^2$ fitting was performed in Sherpa using a simple absorbed power law model.  The column density was fixed at the Galactic value of $3.56\times10^{20}$~cm$^{-2}$ \citep{1990ARA&A..28..215D}.  The individual fits were all acceptable and yielded consistent results, so joint fits were performed with the power law indices tied to each other and the normalizations allowed to float.  The best fit photon index for ObsID 363 is $1.57\pm0.04$ and for ObsID 1630 is $1.54\pm0.07$, which compares well with the values found from the fits of image C alone ($1.55\pm0.09$ and $1.46\pm0.08$, respectively).  Based on the individually fitted power laws, the unabsorbed 0.5--8 keV flux of image C is $(6.2\pm0.4)\times10^{-14}$~\ergcms\ in ObsID 363 and $(6.9\pm0.9)\times10^{-14}$~\ergcms\ in ObsID 1630.  These serve as useful reference fluxes since image C is fairly uncontaminated by flux from the other images and is also a minimum image and therefore less susceptible to fluctuations.

\begin{figure}
\centering
\includegraphics[width=0.47\textwidth]{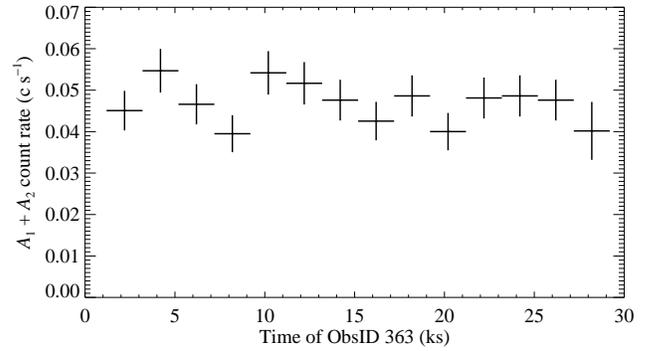}
\caption{Light curve of the 0.5--8 keV count rate of $A_1+A_2$ in ObsID 363 showing a rather constant flux. Horizontal bars indicate the 2 ks time bins, and vertical bars show 1-$\sigma$ errors.}
\label{fig:lc363}
\end{figure}

{\tt ACIS Extract} was also used to obtain light curves from the above extraction regions for each observation.  No significant signs of short-term variability were found within either observation; Kolmogorov-Smirnov tests showed that each light curve had a greater than 10\% chance of being consistent with a constant count rate.  The light curve for the $A_1+A_2$ region is plotted in Figure~\ref{fig:lc363}.  

Given the time delays among the lensed images, it is fair to ask if intrinsic short-term quasar variability combined with a time delay could masquerade as a genuine X-ray flux ratio anomaly.  We can rule this out in the X-ray band for ObsID 363.  The time delay between $A_1$ and $A_2$ from our lens model (see \S\ref{sec:model}) is $14.5\pm2$ ks (with $A_1$ leading).  The 26.5 ks observation therefore covers 1.8 time delay cycles. If we split the observation into two equal parts, we obtain the same $A_1/A_2$ ratio as in Table~\ref{tab:xrayfits}.  To produce this ratio as well as the constant $A_1+A_2$ lightcurve in Figure~\ref{fig:lc363} purely by variability is highly implausible. 

\subsection{Optical observations}
\label{sec:opt-obs}

\pg\ has been observed repeatedly with the Magellan 6.5-meter Baade and Clay telescopes at Las Campanas Observatory between 2001 March and 2006 February using the Raymond and Beverly Sackler Magellan Instant Camera (MagIC).  The instrument has a scale of 0\farcs{}0691 per pixel and a 2.36 arcminute field.  We present here results from three epochs for which the seeing was especially good, making the decomposition of $A_1$ and $A_2$ easier and less uncertain, and reducing the contamination from the lensing galaxy.  Three 60-second exposures were obtained with a Johnson V filter on UT 2001 March 26.  Two 60-second exposures each were obtained obtained with a Sloan $i'$ filter on UT 2004 Feb 22 and 2005 June 07.

The data were flattened using standard procedures.  {\tt ClumpFit}, an empirical PSF-fitting photometry program based on {\tt DoPHOT}, was used to measure fluxes and positions for the four quasar images and for the lensing galaxy.  The profile for the galaxy was taken to be an elliptical pseudo-Gaussian.  As we presently concern ourselves only with flux ratios, we have not put our photometry onto a standard system.  The fluxes for the $A_1$, $A_2$ and $B$ images are given relative to the $C$ image, for which the microlensing fluctuations are expected to be smallest.  It should be remembered that variations of 0.1 mag have been seen on a timescale of weeks and that image $C$ leads the $A$ images and the $B$ image by 10 and 25 days, respectively \citep{1997ApJ...475L..85S, 1997ApJ...489...21B}. The results of our photometry are given in Table \ref{tab:opthistory}, along with selected results (typically those obtained in the best seeing) from prior epochs.

We note that the flux ratios for contemporaneous observations appear to be consistent to within a few percent over the optical wavelength region.  We therefore make no attempt to account for bandpass in presenting the present and past optical results.  

\begin{deluxetable*}{lcccccc}
\tablewidth{0pt}
\tablecaption{Optical Magnitude Differences and Flux Ratios of \pg
\label{tab:opthistory}}
\tablehead{
 & & & \multicolumn{3}{c}{Magnitude differences} & Flux ratios \\
\colhead{UT date} & \colhead{Filter} & \colhead{FWHM} 
& \colhead{$A_1-C$} & \colhead{$A_2-C$} & \colhead{$B-C$} 
& \colhead{$A_2/A_1$}}
\startdata
1984 Mar 26\tablenotemark{a} & B & 0\farcs75 & $-1.26$ & $-1.21$ & 0.41 & $0.95 \pm 0.07$ \\
1985 Mar 19\tablenotemark{a} & V & 0\farcs62 & $-1.18$ & $-0.83$ & 0.49 & $0.73 \pm 0.04$ \\
1986 Feb 19\tablenotemark{b} & V & 0\farcs6  & $-1.27$ & $-0.99$ & 0.48 & $0.77 \pm 0.03$ \\
1986 Feb 19\tablenotemark{b} & B & 0\farcs6  & $-1.23$ & $-0.97$ & 0.48 & $0.79 \pm 0.03$ \\
1991 Mar 03\tablenotemark{c} & F785LP & \hst   & $-1.46$ & $-1.07$ & 0.50 & $0.70 \pm 0.01$ \\
1991 Mar 03\tablenotemark{c} & F555W  & \hst   & $-1.47$ & $-1.02$ & 0.50 & $0.66 \pm 0.01$ \\
1995 Dec 20\tablenotemark{d,e} & V & 0\farcs85    & $-1.50$ & $-1.04$ & 0.47 & $0.66 \pm 0.01$ \\ 
2001 Mar 26\tablenotemark{e} & V & 0\farcs56 & $-1.48$ & $-1.04$ & 0.42 & $0.68 \pm 0.01$ \\ 
2004 Feb 22\tablenotemark{e} & $i'$ & 0\farcs48 & $-1.40$ & $-1.18$ & 0.42 & $0.81 \pm 0.01$ \\
2005 Jun 07\tablenotemark{e} & $i'$ & 0\farcs43 & $-1.40$ & $-1.19$ & 0.42 & $0.81 \pm 0.01$ \\[4pt] 
\colrule
\\
Lens Model & \nodata & \nodata & $-1.48$ & $-1.43$ & 0.44 & 0.96 \\
\enddata
\tablenotetext{a}{Vanderriest et al. (1986)}
\tablenotetext{b}{\citet{1987ApJ...312...45C}}
\tablenotetext{c}{\citet{1993AJ....106.1330K}}
\tablenotetext{d}{\citet{1997ApJ...475L..85S}}
\tablenotetext{e}{present work}
\end{deluxetable*}


\section{Discussion}
\label{sec:discuss}

\subsection{Modeling the lens}
\label{sec:model}

Using Keeton's (2001) {\tt Lensmodel} software, we modeled the lensing potential as a singular isothermal sphere accompanied by a second, offset singular isothermal sphere, which provides a quadrupole moment. This choice of model was motivated by the presence of a group of galaxies to the southwest of the lensing galaxy. We used the image positions provided by the CASTLES Lens Survey\footnote{http://www.cfa.harvard.edu/castles/}, and did not constrain the fluxes. Our best-fit model predicts an Einstein radius of 1\farcs{}0 for the primary lensing galaxy, with a second mass having an Einstein radius of 2\farcs{}6 located 12\farcs{}5 away at a position angle $116^{\mathrm o}$ west of north. This places it close to the observed location of the associated group of galaxies. The model yields a total reduced $\chi^2$ of 3, with the greatest contribution coming from the position of the primary lensing galaxy. The flux ratios predicted by this model are listed in Tables~\ref{tab:xrayfits} and \ref{tab:opthistory}, and may be expected to vary between different plausible models of the lens at the 10\% level. 

\subsection{Anomalous flux ratios and microlensing}
\label{sec:microlensing}

Simple smooth analytic models \citep{2002ApJ...567L...5M} predict that the $A_2/A_1$ flux ratio should be very nearly equal to unity.  For our lens model, the ratio is 0.96. \citet{2005ApJ...627...53C} observe a mid-infrared flux ratio of $0.93 \pm 0.06$, consistent with this prediction.  In 1984, \citet{1986A&A...158L...5V} measured a flux ratio of $0.95 \pm 0.07$, but since then, as seen in Table \ref{tab:opthistory}, the optical flux ratio has varied on a timescale of years between 0.66 and 0.81.  As noted in \S\ref{sec:xray-obs}, the contemporaneous X-ray flux ratio is less than 0.2, inconsistent not only with the predictions of the smooth models, but with the optical observations as well.  

Microlensing by stars in the lensing galaxy could in principle account for such flux ratios, but only if the source is small compared to the Einstein radii of the microlensing stars.  Our simple model has convergence, $\kappa$, and shear, $\gamma$, roughly equal at the image positions, with magnifications $\mu$ of 19.9 for the $A_1$ image and $-$19.0 for the $A_2$ image.  Examples of point source magnification histograms for pairs of images very much like those in \pg\ are presented by Schechter and Wambsganss, with magnifications for $A_1$ and $A_2$ of 10 and 16, respectively \citep{2002ApJ...580..685S}.  They present histograms both for the case when 100\% of the convergence is due to stars and for the case when only 20\% of the convergence is due to stars and the rest is due to a smooth dark component.  The X-ray flux ratio rules out neither hypothesis but is considerably more likely if dark matter is present.

Until now, it was a bit of a puzzle why the optical flux anomalies had failed to deviate from unity as much as was predicted by these histograms. Now it appears that it was because the optical region is too large to be strongly microlensed (see \S\ref{sec:sizes}). As Schechter \& Wambsganss note, the determination of the dark matter fraction of lensing galaxies using the statistics of flux ratio anomalies is made considerably more difficult if the source size is comparable to that of a stellar Einstein ring. It seems now that the X-ray flux ratio anomalies offer a cleaner determination of the dark matter fraction than the optical anomalies.

\subsection{Long-term X-ray variability}
\label{sec:xrayvar}

According to the microlensing model for flux-ratio anomalies, discussed below, $A_2$ is expected to brighten in X-rays on a timescale of $\sim$10 years, and follow-up \chandra\ observations will be able to directly test this. As $A_2$ brightens, the unresolved flux will also increase.  To look for past signs of this effect, we searched the High Energy Astrophysics Science Archive Research Center, provided by NASA's Goddard Space Flight Center, for other X-ray observations of \pg\ and found two \rosat\ observations and three relevant \xmmn\ observations.  The \rosat\ observations and an earlier {\it Einstein} observation are analyzed in \citet{2000ApJ...531...81C}.

\begin{figure}
\centering
\includegraphics[width=0.47\textwidth]{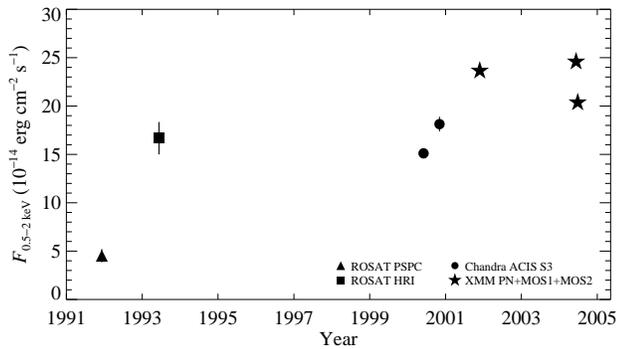}
\caption{Long-term X-ray light curve of \pg\ showing the combined flux of all four images. For most observations, the plotted error bars are smaller than the plotting symbols.}
\label{fig:ltlc}
\end{figure}

The \rosat\ count rates were converted to unabsorbed 0.5--2 keV fluxes using WebPIMMS \citep{1993Legac...3...21M} with the assumptions of an absorbed power law of photon index 1.65 and a column density of $3.56\times10^{20}$~cm$^{-2}$.  For the \xmm\ observations, we extracted spectra of \pg\ from the EPIC-PN and both EPIC-MOS detectors.  We performed joint spectral fits (on all quasar images added together) in the 0.5--10 keV band for each observation with simple absorbed power laws with the column density fixed at the Galactic value.  These gave acceptable fits, from which we computed the unabsorbed 0.5--2 keV fluxes.  We also used our previous \chandra\ joint fits to compute the total 0.5--2 keV fluxes (from all quasar images added together) from the \chandra\ observations.  The long-term X-ray light curve is shown in Figure~\ref{fig:ltlc}.

From the seven measurements of the lensed flux from \pg\ over the course of 12.5 years, the mean is $1.75\times10^{-13}$~\ergcms, and the sample standard deviation is $6.7\times10^{-14}$~\ergcms, or $\sim$40\%.  There is no evidence for strong short term variability from the individual lensed images in the \chandra\ data, nor is there evidence for strong short term variability within the three \xmm\ observations (in which the individual images are unresolved). 

As discussed above, if the demagnification of $A_2$ is due to microlensing,  the unresolved flux will rise as $A_2$ becomes less demagnified.  The observed relative X-ray fluxes of the four images $A_1:A_2:B:C$ are $1:0.16:0.25:0.25$ (based on ObsID 363; see Table~\ref{tab:xrayfits}).  If $A_2$ were to rise in flux to match $A_1$, the overall change in flux would be $\sim$50\%.  The recent \xmm\ observations show that the X-ray flux has risen $\sim$30\% since the \chandra\ observations from six years ago (Figure~\ref{fig:ltlc}).  However, there is an obvious degeneracy between a rise in the flux of $A_2$ and typical quasar variability over the course of many years.

\subsection{Sizes of quasar emission regions}
\label{sec:sizes}

The size scales of the emission regions in quasars are difficult to probe directly since they are on the microarcsecond scale or smaller. The use of temporal variability for inferring sizes is indirect and becomes impractical for distant quasars.  By contrast, microlensing directly explores angular scales of (by definition) microarcseconds. Of the emission features of the quasar, only those which subtend smaller angles on the sky than the Einstein radius of the microlenses will exhibit strong variations in flux.

\begin{figure}
\centering
\includegraphics[width=0.47\textwidth]{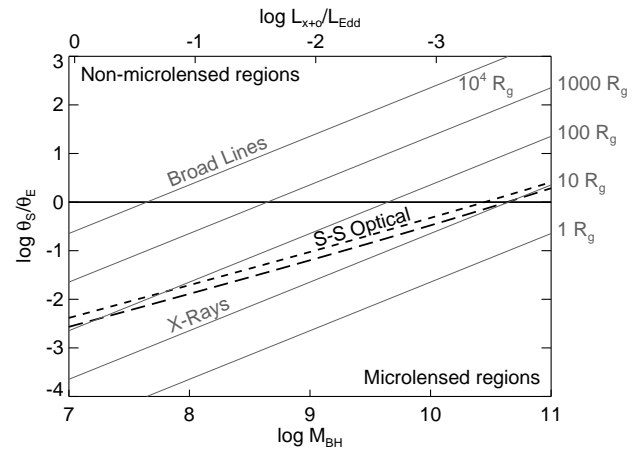}
\caption{Source sizes at X-ray and optical wavelengths (see text).}
\label{fig:fxi}
\end{figure}

Figure~\ref{fig:fxi} displays the results from our study of \pg.  Here we have plotted the ratio of angular scale for different regions of the quasar, $\theta_s$, to the Einstein radius of a solar-mass microlens, $\theta_E$.  For ratios greater than unity, microlensing should be strongly suppressed \citep[for a detailed analysis see][]{2005ApJ...628..594M}.  The ratio $\theta_s/\theta_E$ is plotted against the assumed mass of the central black hole, $M_{\rm BH}$. For every value of $M_{\rm BH}$ there is a corresponding Eddington luminosity which can be compared to the observed values of \Lx\ ($2.4\times 10^{44}$~\ergcms; 0.5--8 keV; this work) and $L_{\rm opt}$ ($1.2\times 10^{45}$~\ergcms, from a sum of the V-, I-, and H-band data provided by the CASTLES Lens Survey) for \pg\ (see the top axis label).  Within the $\theta_s/\theta_E$ vs.\ $M_{\rm BH}$ plane we plot contours of constant size in units of $R_g$, the gravitational radius of the black hole ($GM_{\rm BH}/c^2$).  As is evident from the plot, the X-rays, which should arise deep in the gravitational potential well of the black hole, should be microlensable for any $M_{\rm BH} \lesssim 10^{10}\,M_\odot$.  This is in clear agreement with the large X-ray flux ratio anomalies observed for \pg\ and for two other quad lenses: RX\,J0911+0554 and RX\,J1131$-$1231 \citep{2001ApJ...555....1M,2006ApJ...640..569B}.  By contrast, the broad-line emission region should not be microlensable, except for a lower mass black hole (i.e., $M_{\rm BH} \lesssim 3 \times 10^7\,M_\odot$).  Finally, the dotted and dashed curves mark the radii within which 50\% of the power in the $I$ and $V$ bands emerge, respectively, for a simple thin accretion disk model \citep[e.g.][]{1973A&A....24..337S}.  According to these curves, the optical continuum ought to be microlensed by approximately the same amount as in the X-ray band, in agreement with \citet{2005ApJ...628..594M}. But clearly it is not!

Using \hst\ spectra, \citet{2005MNRAS.357..135P} found that the $A_2/A_1$ ratio in the ultraviolet continuum is $\sim$0.5 and decreases to shorter wavelengths, indicating that the UV is more severely microlensed than the optical but less microlensed than the X-rays.

Therefore, within the microlensing scenario, we can conclude that the continuum optical emission from \pg\ comes from much further out than the UV, which in turn comes from further out than the X-rays.  In particular, we find that the optical emission comes from a region $\sim$10--100 times larger than expected for a thin accretion disk model (for $M_{\rm BH}$ in the range $3 \times 10^9\,\rightarrow 10^8\,M_\odot$). Since $L_{\rm opt}$ dominates \Lx\ in \pg\ (and for many other luminous quasars), this is difficult to understand from an energetics point of view, since the energy released goes as $r^{-1}$.  Of course the optical light could be scattered by a large-scale plasma region; however, in that case one would expect the X-rays to be scattered as well, and hence share a similar effective emission region.  Thus, while the X-ray images clearly appear to be microlensed, the bulk of the optical emission must be coming from $\sim$100--3000\,$R_g$ from the central black hole (for $M_{\rm BH}$ in the range $3 \times 10^9\,\rightarrow 10^8\,M_\odot$).

In coming to these conclusions, we have neglected special- and general-relativistic effects in the emissions from the accretion disk, except for cosmological redshift. In addition, we have followed \citet{2005ApJ...628..594M} in assuming a Kerr black hole with a large spin parameter ($a=0.88$). This is consistent with estimates for a typical quasar \citep{2006astro.ph..3813W}, and implies an innermost disk radius of $2.5 R_g$ and a binding energy per mass $\eta=0.146$. We have also set the bolometric luminosity to 33\% of the Eddington luminosity, as advocated by \citet{2005astro.ph..8657K}. Neither of these parameter assumptions has a strong effect on the size of the predicted optical emission region for a thin accretion disk model.

\bigskip
\section{Conclusions}
\label{sec:conclusions}

We have made use of optical data collected over the past 22 years to demonstrate that the bright, close pair of lensed images of \pg\ has a consistent flux ratio ($A_2/A_1$) of $\sim$0.7--0.8.  X-ray observations with \chandra, covering two epochs separated by 5 months, indicate a much more extreme flux ratio of $\sim$0.2.  Both the optical and X-ray ratios are anomalous with respect to smooth lensing models, which predict a flux ratio of 0.96.  We used a comparison of the optical and X-ray flux ratio anomalies to argue in favor of the microlensing origin of the anomalies, and to show that the optical emission region is much larger (i.e., $\sim$$10-100$) than predicted by a simple thin accretion disk model.   

\acknowledgements 

We thank Joachim Wambsganss, George Chartas, and the anonymous referee for useful comments.  DP gratefully acknowledges support provided by NASA through Chandra Postdoctoral Fellowship grant number PF4-50035 awarded by the Chandra X-ray Center, which is operated by the Smithsonian Astrophysical Observatory for NASA under contract NAS8-03060.  SR received some support from Chandra Grant TM5-6003X. JAB and PLS acknowledge support from NSF Grant AST-0206010. This research has made use of data obtained from the High Energy Astrophysics Science Archive Research Center (HEASARC), provided by NASA's Goddard Space Flight Center.

\end{document}